\documentclass[useAMS,usenatbib]{mn2e}
\usepackage{psfig}
\usepackage{graphicx}

\newcommand{\be}{\begin{equation}}
\newcommand{\ee}{\end{equation}}
\newcommand{\bea}{\begin{eqnarray}}
\newcommand{\eea}{\end{eqnarray}}
\newcommand{\rd}{{\rm d}}

\title[Cross-correlation of 2MASS and WMAP3]{Cross-correlation of 2MASS and WMAP3: 
Implications for the Integrated Sachs-Wolfe effect}
\author[A. Rassat, K. Land, O. Lahav, F.B. Abdalla]{Ana\"\i s
Rassat$^{1}$\thanks{E-mail: ammr@star.ucl.ac.uk}, Kate
Land$^{2}$\thanks{E-mail: kate.land@ic.ac.uk},
Ofer Lahav$^{1}$\thanks{E-mail: lahav@star.ucl.ac.uk}
Filipe B. Abdalla$^{1}$\\
$^{1}$Department of Physics and Astronomy, University College London, Gower Street, London, WC1E 6BT\\
$^{2}$Theoretical Physics, Imperial College, Prince Consort Road,
London, SW7 2AZ}

\begin{document}

\date{Accepted xxx. Received xxx; in original form xxx}

\pagerange{\pageref{firstpage}--\pageref{lastpage}} \pubyear{2006}

\maketitle

\label{firstpage}

\begin{abstract}
We perform a cross-correlation of the Cosmic Microwave Background
(CMB) using the third year Wilkinson Microwave Anisotropy Probe 
(WMAP) data with the 2 Micron All Sky Survey (2MASS) galaxy
map (about 828 000 galaxies with median redshift $z\approx 0.07$). 
One motivation is to detect the
Integrated Sachs-Wolfe (ISW) effect, expected if the cosmic gravitational
potential is time dependent; for example, as it is in a flat universe with a
Dark Energy component. The measured spherical harmonic
cross-correlation signal favours the ISW signal expected in the
concordance $\Lambda$CDM model over that of zero correlation, although
both are consistent with the data within 2$\sigma$. Within a flat $\Lambda$CDM model
we find a best fit value of $\Omega_\Lambda=0.85$ and
$\Omega_{\Lambda} < 0.89$ (95\% CL). The above limits assume a
galaxy bias $b_{\rm g}\big(\frac{\sigma_8}{0.75} \big)\approx 1.40 \pm 0.03$, 
which we derived directly from
the 2MASS auto-correlation. Another goal is to test if previously reported
anomalies in the WMAP data are
related to the galaxy distribution (the so-called ``Axis of
Evil'' - AoE). No such anomaly is observed
in the 2MASS data nor are there any observed AoE correlations
between the 2MASS and
WMAP3 data.

\end{abstract}

\begin{keywords}
Dark Energy -- Cosmic Microwave Background -- Integrated
Sachs-Wolfe effect
\end{keywords}

\section{Introduction}\label{intro}

Our currently favoured $\Lambda$CDM cosmological model has
received on-going confirmation and bolstering over recent years,
especially from recent observations of the cosmic microwave
background (CMB) by the Wilkinson Microwave Anisotropy Probe
(WMAP)~\citep{Spergel:2003cb,Spergel:2006hy}. This model
postulates that more than two thirds of the Universe is composed
of `Dark Energy', a mysterious energy with negative pressure.
This Dark Energy has never been directly observed - only
inferred. The case is compelling because of the very different
types of observations that require its presence: acceleration of the
Universe seen by supernov{\ae}; joint analysis of the CMB with Large Scale Structure (LSS) requiring
zero curvature~\citep{Hinshaw:2006ia, Spergel:2006hy}
but a mass density today $\Omega_{\rm m}\sim0.25$. However, it would still be comforting to
have a more direct and independent detection of the Dark Energy
and its effects, especially in light of promising alternative
theories based on
inhomogeneous models which do not require Dark Energy~\citep{Alnes:2005rw,Moffat:2006ug,Vanderveld:2006rb}.

The Integrated Sachs-Wolfe (ISW) effect~\citep{Sachs:1967er}
provides us with one of the cleanest probes of Dark Energy,
though it does require some assumptions about other cosmological parameters. As CMB photons travel through
space they pass through the gravitational potential wells of LSS. As they fall into a well, the photons are
blueshifted, and as they climb out they are redshifted. In an
Einstein-de Sitter universe, these effects cancel and no net shift is
observed on large scales. However, if Dark Energy dominates, large scale
gravitational potentials decay and there is an overall net effect.
This secondary anisotropy is called the late-time ISW effect.

On large scales, the ISW effect will add power to the CMB anisotropies, by:

\be \Big(\frac{\Delta T}{T}\Big)_{ISW}=
-2\int_{\eta_{\rm L}}^{\eta_0}\Phi'\big((\eta_0-\eta) \rm{\bf{\hat n}},\eta \big)\rd \eta \ee
 where $T$ is the temperature; $\eta$ is the conformal time, defined
 by $\rd \eta = \frac{\rd t}{a(t)}$, and $\eta_0$ and $\eta_{\rm L}$ are
 the conformal times today and at the surface of last scattering
 respectively; $\rm{\bf{\hat n}}$ is the unit
 vector along the line of sight; $\Phi(\bf{x},\eta)$ is the gravitational potential at
 position $\bf{x}$ and at conformal time $\eta$, and $\Phi'
 \equiv \frac{\partial \Phi}{\partial \eta}$.  Its relative amplitude makes it difficult
to distinguish from the primary anisotropies.  However,~\cite{Crittenden:1995ak} proposed using the
cross-correlation between the LSS and the CMB to
detect the ISW effect, independently from the intrinsic CMB fluctuations. 
In the case where the gravitational potentials decay, a positive correlation is expected. 
This means that on large scales hot spots in the
CMB will correspond to overdense regions in the galaxy distribution.
This positive correlation is also expected in open
 universes~\citep{Kamionkowski:1996ra, Kinkhabwala:1999k}, whereas a negative correlation will
occur in closed universes.  The Sunyaev-Zeldovich effect also produces a negative correlation,
but this is expected on smaller scales ($\ell>20$ for $z \approx 0.07$,
 see ~\cite{Afshordi:2003xu}).

In our currently favoured cosmological models, we believe the Universe has recently ($z<1$) become
dominated by Dark Energy, which makes the ISW effect a fitting probe of our
cosmological model, though alternative models of
gravity predict similar signatures ~\citep{Carroll:2004de, Song:2006sh}.

The first cross-correlations between the COBE (Cosmic Microwave
Background Explorer) CMB map and tracers of the LSS (hard X-ray
background and Radio sources) did not reveal any significant
detections~\citep{Boughn:2002, Boughn:2003}; however, there have
since been a number of reported detections of late-time ISW from
cross-correlating the first-year WMAP data (WMAP1) with: Radio
sources~\citep{Nolta:2003uy, Boughn:2003yz, Boughn:2004zm}; the hard X-ray
background~\citep{Boughn:2003yz,Boughn:2004zm}; the Sloan Digital
Sky Survey
(SDSS)~\citep{Scranton:2003in,Fosalba:2003ge,Padmanabhan:2004fy};
the 2 Micron All Sky Survey (2MASS)~\citep{Afshordi:2003xu}; the
APM Galaxy Survey~\citep{Fosalba:2004ge}; a combination of the
above~\citep{Gaztanaga:2004sk}.  Recently, the third-year WMAP data
(WMAP3) was correlated by ~\cite{Cabre:2006qm} with the fourth SDSS
data release (DR4), and they detected a significant positive
cross-correlation, while ~\cite{Giannantonio:2006al} cross-correlated
it with high redshift SDSS quasars and found a
~2$\sigma$ detection. The above cross-correlations were all performed
in angular or harmonic space.~\cite{McEwen:2006md} used a
directional spherical wavelet analysis and found a positive
detection at the 3.9$\sigma$ level.

A cross-correlation between LSS and the
CMB can also probe the issue of foreground contamination in the
CMB and LSS maps. Recent studies have highlighted anomalous
features in CMB data that include alignments of the low-$\ell$
multipoles, and a North-South
asymmetry~\citep{Land:2005ad,Land:2004bs,Copi:2005ff,Eriksen:2003db}.
Curiously these features align with the ecliptic plane, and possibly 
the Supergalactic Plane. Such alignments may, of course, be coincidental, 
or they may indicate the culprit behind these
anomalies, such as contamination or some other more local secondary effect of
the structure on CMB photons~\citep{Vale:2005mt,Rakic:2006tp}.

2MASS as an infrared survey has the advantage of probing the older
stellar populations in galaxies and is therefore a robust tracer of
their mass.  Since infrared light penetrates more easily through the
Galaxy than optical light, 2MASS is nearly an all-sky extra-galactic
survey.
These two qualities of 2MASS make it one of the best available tracers of any
large scale correlations, such as those expected by the ISW effect. 
However, even in the infrared, Galactic extinction and stellar
contamination will produce a foreground contamination in the galaxy distribution which can be
correlated with emission foregrounds that contaminate WMAP.

In this paper we perform a harmonic space cross-correlation of the
CMB with LSS using the WMAP3 temperature
maps~\citep{Spergel:2006hy, Hinshaw:2006ia} and the 2MASS galaxy survey
~\citep{Jarrett:2000me, Jarrett:2004a}. We account for possible
correlations from Galactic contamination using random simulations.  This work updates
the work of~\cite{Afshordi:2003xu}, who cross-correlated  WMAP1 with
2MASS.  We also investigate correlations between another statistic,
the so-called ``Axis of Evil'' (AoE) ~\citep{Land:2005ad}.

In Section~\ref{data} we introduce the 2MASS and WMAP3 data that we
use, and in Section~\ref{theory} we outline our theory.  In
Section~\ref{priors} we discuss our choice of fiducial model.  In
Section~\ref{bias} we estimate the galaxy bias, $b_{\rm g}$, from the galaxy
auto-correlation function. In Section~\ref{method} we discuss our cross-correlation
method. In Section~\ref{results} we
present our results of the cross-correlation and discuss their statistical significance.  In Section~\ref{axis} we
investigate AoE type correlations and in
Section~\ref{discuss} we present a discussion of our results.

\section{The Data}\label{data}

In this section we summarize the two data sets used for the
cross-correlation.

\subsection{The Large-Scale Structure: 2MASS}
We use the publicly available full-sky extended source catalog
(XSC) of the near-infrared 2MASS~\citep{Jarrett:2000me}. Following~\cite{Afshordi:2003xu},
we divide the galaxies into different magnitude bands depending on
their $K_s$-band isophotal magnitude $K_{20}$ (``k\_m\_i20c'' -
20mag/arcsec$^2$ isophotal circular ap. magnitude). We correct
these magnitudes for Galactic extinction using the IR reddening
maps of~\cite{Schlegel:1997yv}\footnote{Corresponding reddening maps can be found
at http://astro.berkeley.edu/davis/dust/index.html}: \be K_{20} \to
K_{20}-A_K\ee where $A_K=0.367\times
E(B-V)$~\citep{Afshordi:2003xu}.  We use the same extinction maps
to create a mask that excludes regions of sky where the XSC is
unreliable.~\cite{Afshordi:2003xu} find a limit of $A_k<0.05$ for
which 2MASS is seen to be 98\% complete for $K_{20}<13.85$. We
adopt this method herein, and mask areas with $A_k>0.05$, leaving
69\% of the sky with approximately 828,000 galaxies for the analysis.

\begin{figure}\begin{center}\leavevmode
\psfig{file=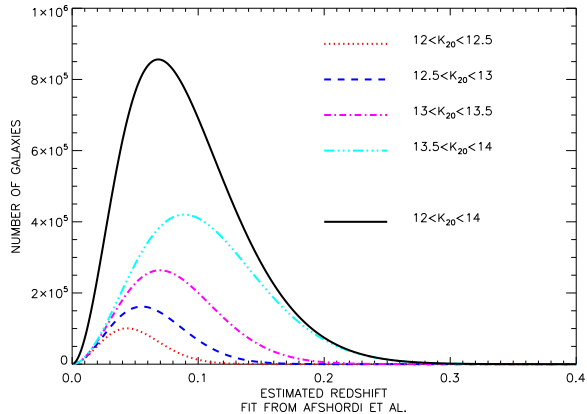,width=8cm}
\end{center} \caption{The redshift distribution for different
    magnitude cuts of 2MASS,
estimated from the parametrization in Afshordi et al
(2004).  The red (dotted) line corresponds to the closest shell
$12<K_{20}<12.5$, the blue (dashed) to the $12.5<K_{20}<13$ shell,
the magenta (dot-dashed) to the $13<K_{20}<13.5$ shell, the cyan
(dot-dot-dashed) to the furthest $13.5<K_{20}<14$ shell.  The solid
(black) shows all four shells combined, i.e. $12<K_{20}<14$.  The
redshift, $z_0$, at which each distribution peaks and the number
counts for the four $K_{20}$ shells are reported in Table~\ref{dNdz}.}\label{figdNdz}
\end{figure}

\begin{figure*}
\centerline{\psfig{file=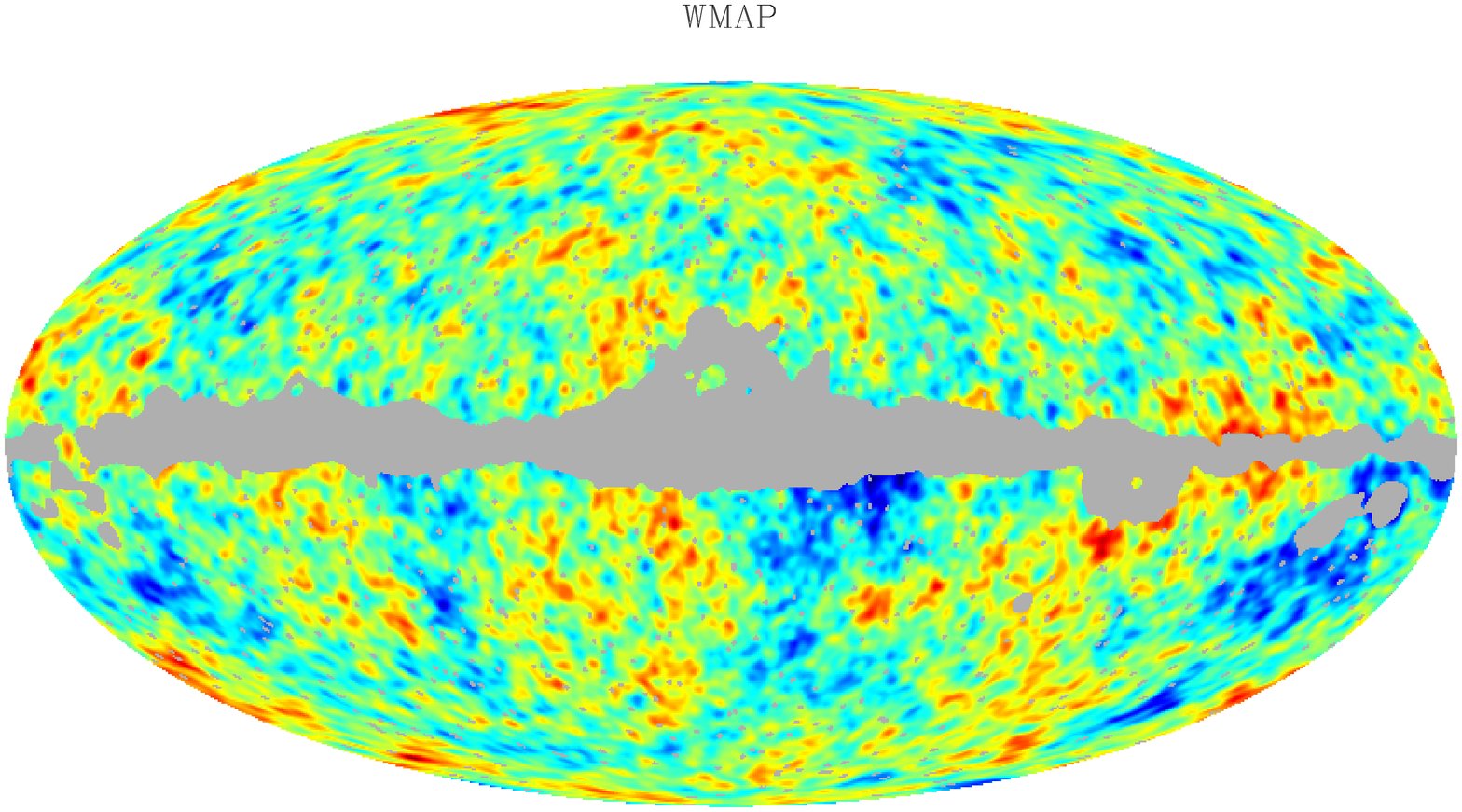,width=15.0cm,angle=90}}
\centerline{\psfig{file=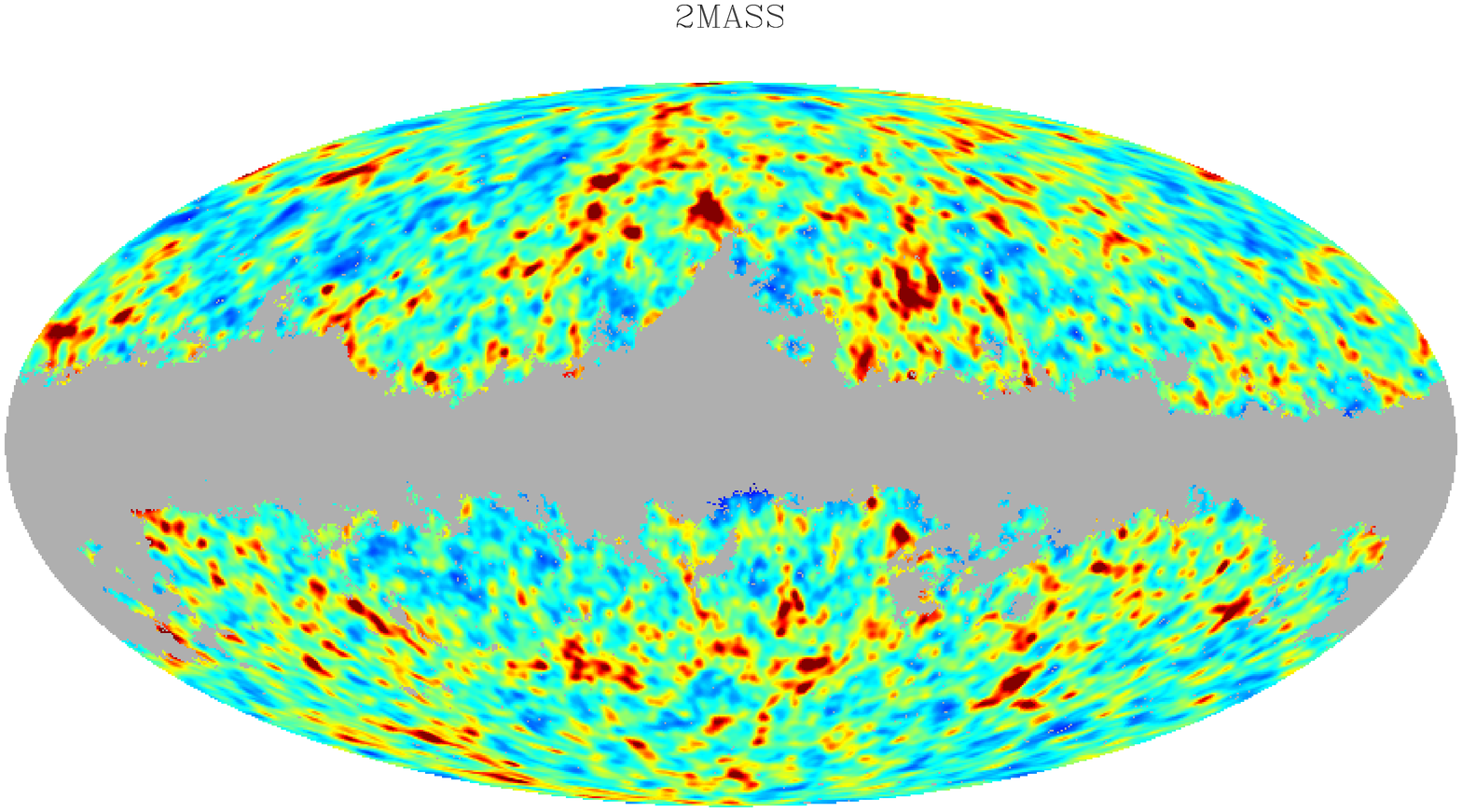,width=15.0cm,angle=90}}
\caption{ {\bf Best viewed in colour.}\emph{Top:} The 3rd year WMAP
(WMAP3) Internal Linear Combination (ILC) map, which maps temperature anisotropies in the Cosmic
Microwave Background.  \emph{Bottom:} The 2 Micron All-Sky Survey
(2MASS) overdensity field of 827,947 galaxies with magnitudes
$12<K_{20}<14$.  Red shading represents CMB hot spots and 2MASS
overdensities and blue shading corresponds to CMB cold spots and 2MASS
underdensities respectively.  Grey shading corresponds to the $K_p2$ mask for the
CMB and for the 2MASS map to galaxies with galactic extinction $A_k>0.05$.  
Both maps are produced with Healpix, are in Mollweide
projection of Galactic Coordinates and are convolved with a gaussian
beam of FWHM 100'.  The center of the projection represents galactic
coordinates $(l, b)=(0,0)$ and the longitudinal coordinate increases
leftwards and $l=180^\circ$ on both left and right edges of the projection.  
The data are publicly available from
http://lambda.gsfc.nasa.gov/ (WMAP3, see Hinshaw et al. 2006) and
http://www.ipac.caltech.edu/2mass/ (2MASS, see Jarrett 2004).}\label{figdata}
\end{figure*}

For our four $K_{20}$ magnitude shells, we adopt the redshift distribution
computed by~\cite{Afshordi:2003xu} (parametrized by their
Equations 33 and 35). They fit the redshift distribution from
the 2MASS $K_{20}$ luminosity function~\citep{Kochanek:2000im} for
a three parameter generalized gamma distribution which we use
herein. We recall the number counts for the four $K_{20}$ shells  and the redshift $z_{0}$ at which the
distributions peak, taken from~\cite{Afshordi:2003xu}, in
Table~\ref{dNdz}.  In Figure~\ref{figdNdz} we plot the redshift distribution for each shell
as well as for $12< K_{20}< 14$.  The 2MASS overdensity field for
galaxies with galactic extinction $A_k<0.05$ is plotted in
Figures~\ref{figdata}, where it is convolved with a gaussian beam of
Full Width Half Maximum
(FWHM) 100'.

\begin{table}
\caption{The number of galaxies for our four $K_{20}$ shells and for
  all four shells combined. We
also recall the peak redshift values for the
distributions.  These values are taken from Afshordi et al. (2004).  The parametrizations of these distributions are shown
  in Figure~\ref{figdNdz}.}

 \centering
  \begin{tabular}{lcc}
  \hline
Magnitude & $N_{tot}$ & $z_{0}$\\
 \hline
$12.0 < K_{20} < 12.5$ & 49902  & 0.043 \\
$12.5 < K_{20} < 13.0$ & 102947 & 0.054 \\
$13.0 < K_{20} < 13.5$ & 217831 & 0.067 \\
$13.5 < K_{20} < 14.0$ & 457267 & 0.084 \\
 \\
$12.0 < K_{20} < 14.0$ & 827947 & 0.073 \\

\hline
\end{tabular}\label{dNdz}
\end{table}

We make maps of the 2MASS overdensity using the Healpix
format~\citep{Gorski:1998vw,Gorski:2004by}, and we use their `{\bf
map2alm}' routine to obtain the spherical harmonic coefficients.
For an incomplete sky, the magnitude of the angular power spectrum scales with the
survey area and we compensate for the loss of sky cover by including
the factor $f_{sky}$ in Equation \ref{fsky}.  The sky cut will
also induce correlations between adjacent multipoles.  

\subsection{The Cosmic Microwave Background: WMAP3}
We use the third-year data from NASA's WMAP satellite
~\citep{Hinshaw:2006ia,Spergel:2006hy}\footnote{Available at
http://lambda.gsfc.nasa.gov/}. We use the foreground reduced maps
of the Q (41 Ghz),V (61 GHz), and W (94 GHz) bands. The foreground reduced sky maps were
produced by removing a foreground model from the ``unreduced'' maps.
Synchrotron, free-free, and dust emission templates were modelled
and then subtracted from the single year ``unreduced'' maps. Full
three year maps were produced by performing a weighted,
pixel-by-pixel, mean of the three single year maps. This same
weighted mean method was then used to combine three year maps of
the same frequency band into a single map for each frequency. Two
maps were combined to produce the Q and V band maps; four maps
were combined to produce the W band map. We use the Kp2 mask to
exclude the Galactic plane and other known foreground sources. We
also use the WMAP3 ILC map with a Kp2 mask. The WMAP3 ILC map
convolved with a gaussian beam with FWHM 100' is
plotted in the top part of Figure~\ref{figdata} in Mollweide projection.  As with the
2MASS data, we use the Healpix routine `{\bf map2alm}' to find the
spherical harmonic coefficients.

We also return to the WMAP1 data and in
Section~\ref{method} we compare the cross-correlation results for
the V band (which is the
least contaminated band with the best resolution). We use the first-year foreground cleaned
maps, with the Kp2 mask, and inverse-noise coadd the two available
V band maps.

\section{THEORY}\label{theory}

In this section, we present the formalism for the calculation of Auto-
and Cross-Correlation Functions (ACF \& CCF).

 In what
follows, we use $r$ as the comoving distance and implicitly as a label of
redshift epoch $z$.  They are related for a given cosmology by $\rd r
=\frac{c}{H(z)}\rd z$, where $c$ is the speed of light in vacuum. 
Thus, the growth factor $D(z)$, the growth function
$f(z)$ (defined below) and the Hubble parameter $H(z)$ all have
implicit dependences on $r$. 

 For the angular CCF:

\be C_{gT}(\ell)=4\pi b_{\rm g}\int \rd k \:
\frac{\Delta^2(k)}{k}\:W_g(k)W_{T}(k)\label{CgtEx}\ee

For the angular overdensity ACF:

\be C_{gg}(\ell)=4\pi b_{\rm g}^2 \int \rd k \:
\frac{\Delta^2(k)}{k}\:\Big|W_g(k)\Big|^2\label{CggEx}\ee

where:

\bea \Delta^2(k)&=&\frac{4\pi}{(2\pi)^3}k^3P(k) \\
W_g(k)&=&\int \rd r \: \Theta(r) j_{\ell}(kr)D\\
W_T(k)&=&-\frac{3\Omega_{m,0}H_0^2}{k^2c^3}\int_0^{z_{L}}\rd r
\: j_{\ell}(kr)H D\big(f-1\big) \label{fminus1}\\
\Theta(r)&=&\frac{r^2n_c(r)}{\int \rd r\: r^2 n_c(r)}\eea

  Integrals are performed over the wavenumber $k$ (expressed in
$h{\rm Mpc}^{-1}$), and the comoving distance $r$ (in $h^{-1}{\rm Mpc}$).
The matter power spectrum, $P(k)$, is related to the galaxy power
spectrum through the linear bias $b_{\rm g}$ which we take to be
constant on the depth of our survey. This is justified by the \emph{Galaxy conserving model}
of~\cite{Magliocchetti:2000}
(their Equation 2) where the bias evolution between redshift zero and
the mean redshift of 2MASS was found to be less than 1\%.  The
selection function, $n_c(r)$, for each magnitude shell, is shown in Figure~\ref{dNdz}.

The linear growth factor, $D(z)$, is given by:
\be{D(z) \propto H(a)\int^{a(z)}\frac{\rd a'}{(a'H(a'))^3}}\ee  and
is normalized such that
$D(0)=1$.  In Equation \ref{fminus1}, the redshift at the surface of last
scattering is $z_{L}\sim 1089$. The growth function $f$ is given by $f(z) \equiv \frac
{d\ln D(z)}{d\ln a(z)}$ and can be approximated by $f(z) \simeq
\Omega_m^{0.6}(z)$~(\cite{Peebles:1993}, their Equation 5.120). 
The dependence on the Dark Energy density and
equation of state is negligible at the present epoch
~\citep{Lahav:1991lp,Wang:1998s}.
The Hubble constant is parametrized by
$H_0=100h~{\rm km~s}^{-1}~{\rm Mpc}^{-1}$ and $j_{\ell}(kr)$ is the spherical
bessel function of the 1st kind of order $\ell$.

\begin{figure}
\centering \psfig{file=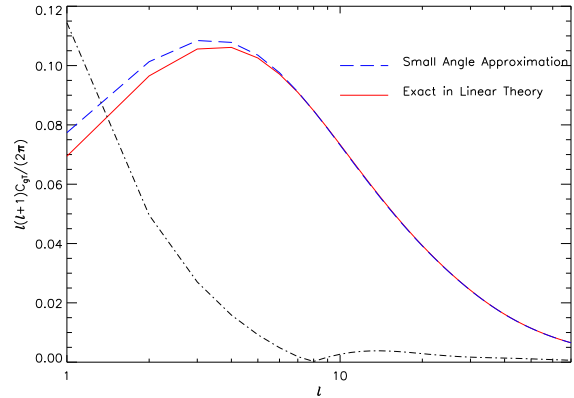,width=8cm}\caption{The
expected cross-correlation from the equation exact in linear
theory (Equation~\ref{CgtEx} - solid red line) and using a large-angle approximation for
the bessel function (Equation~\ref{CgtSA} - dashed blue line).  These values are obtained
using our fiducial model (flat universe with $\Omega_m=0.30$,
$\Omega_b=0.05$, $h=0.7$, $\sigma_8=0.75$) and the selection function
for galaxies with $12<K_{20}<14$.  The black dot-dashed line
is the absolute relative difference between the two. The
difference is less than $1\%$ from $\ell = 5$ onwards, but is
considerably larger for lower multipoles.  For $\ell = 2$, for example, 
the difference is of order 10\%.  The
multipole at which the small angle approximation begins to hold
increases with the mean redshift of the survey.}\label{approx}
\end{figure}

These equations are exact in linear theory, and for $\ell > 10$ we
replace them with the small angle approximation or \emph{Limber}
equation ~\citep{Afshordi:2003xu}, which arise from the bessel
function approximation: \be
\lim_{\ell->\infty}j_\ell(x)=\sqrt{\frac{\pi}{2\ell+1}}\;\delta\big(\ell+\frac{1}{2}-x\big)\ee
Equations~\ref{CgtEx} and~\ref{CggEx} then simplify to:

\be C_{gT}(\ell)=\frac{-3b_{\rm g}H_0^2\Omega_{m,0}}{c^3(\ell+1/2)^2}\int
\rd r\;\Theta D^2 H [f-1]P\left(\frac{\ell+1/2}{r}\right)
\label{CgtSA}\ee

\be C_{gg}(\ell)=b_{\rm g}^2\int \rd
r\:\frac{\Theta^2}{r^2}\:D^2P\left(\frac{\ell+1/2}{r}\right)
\label{CggSA}\ee
Reducing the number of integrals significantly reduces computation 
time. In Figure~\ref{approx} we compare the CCF using the equation
exact in linear theory (Equation ~\ref{CgtEx}) with its small angle
approximation (Equation ~\ref{CgtSA}). The
difference is less than $1\%$ from $\ell = 5$ upwards.  However, for
lower multipoles the difference is more important; for example, at
$\ell = 2$ it is of order 10\%.  The small
angle approximation makes the assumption that $kr \sim (\ell+1/2)$.
From this we see that for deeper surveys, the small angle
approximation will begin to hold at a higher
multipole.  One should therefore check at which multipole the \emph{Limber}
equation begins to hold. Throughout this paper we use Equations~\ref{CgtEx}
and~\ref{CggEx} (\ref{CgtSA} and~\ref{CggSA} when $\ell > 10$) to
model our theory.

\begin{figure}\centering
\psfig{file=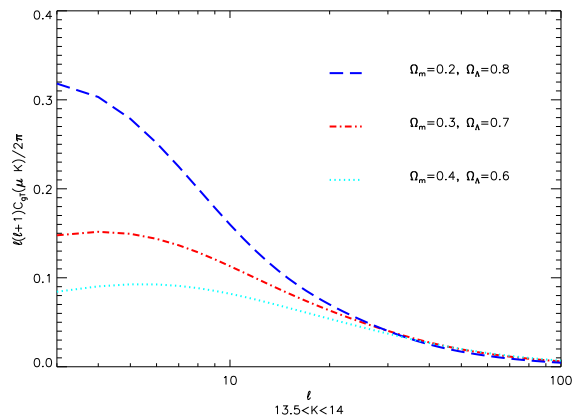,width=8cm}\caption{The expected ISW
  signal ($\Omega_b=0.05$, $h=0.7$, $\sigma_8=0.75$, $b=1.40$) for the
  fourth magnitude bin for flat cosmologies with a cosmological
  constant.  Increasing correlation corresponds to increasing values
  of $\Omega_\Lambda$.  The dotted line corresponds to a cosmology with
  $\Omega_m=0.40$, the dash-dotted line to $\Omega_m=0.30$ and the
  dashed line to $\Omega_m=0.20$.  The ISW effect is strongest at
  scales $\ell < 30$.  An ISW effect is also observed in open cosmologies 
without Dark Energy (Kamionkowski \& Spergel 1994; Kamionkowski 1996; 
Kinkhabwala \& Kamionkowski 1999).}\label{cosmo}
\end{figure}

In Figure~\ref{cosmo} we plot the CCF for different values
of the matter density, $\Omega_m$, for flat cosmologies with Dark
Energy. The correlation increases with
increasing $\Omega_\Lambda$.  We note that an ISW effect is also observed in open
cosmologies \emph{without} Dark Energy~\citep{Kamionkowski:1994s,Kamionkowski:1996ra,Kinkhabwala:1999k};
and for low redshifts ($z<2$) the signal expected in an open cosmology
can be greater than the signal expected in a flat $\Lambda$CDM
cosmology \citep{Kamionkowski:1996ra}.  Closed cosmologies would
produce a negative correlation.  
In fact, within the family of
cosmological models with arbitrary matter density, cosmological
constant and curvature, only the Einstein-de Sitter universe
($\Omega_m=1, \Omega_\Lambda=0$) will give zero correlation.  This can be
seen from the last term in Equation~\ref{fminus1}; for no correlation,
the following condition must be satisfied:$~ f(z) \simeq
\Omega_m^{0.6}(z)= 1$ for all redshift.

\section{Choice of Fiducial Model and Priors}\label{priors}
In this paper we focus on how to constrain Dark Energy
from the ISW effect alone.
This requires of course assumptions about the other cosmological
parameters.
To have the ISW result independent of the CMB,
we prefer to assume ``round'' cosmological parameters which are in accord
with other cosmological measurements, rather than adopting
exact values from another analysis, \emph{e.g.}, the WMAP3 TT correlation
function.
                                                                                                                             
Based on inflation, we assume the Universe is flat,
therefore $\Omega_\Lambda = 1 - \Omega_m$.
We also assume the spectral index to have the Harrison-Zeldovich value
$n=1$ (although some inflationary models and the recent WMAP3 data
suggest $n \approx 0.95$).
Based on the HST key project, we take for the
Hubble parameter $h=0.7$ (the 1$\sigma$ error bar is about 10\%).
For the matter density we take
$\Omega_m = 0.30$, in accord with supernov{\ae}  Ia data combined with the
flatness of the universe (we note however that the recent 2dFGRS data (Cole et al. 2005)
and WMAP3 ~\citep{Spergel:2006hy} favour  $\Omega_m \approx 0.25$).
For the baryon density we assume $\Omega_b = 0.05$, based on
Big Bang nucleosynthesis ~\citep{copi:1995st}.
                                                                                                                             
The normalization of the power spectrum, parametrized as $\sigma_8$, is
still highly uncertain, with reported values in the range
$\sigma_8 \approx  0.75$ (e.g. the recent WMAP3 result) to $\sigma_8 \approx
1.0$ ~\citep{Massey:2005r}.
We fix $\sigma_8 =0.75$, and we solve for the galaxy biasing $b_{\rm g}$.
We note that in linear theory we actually constrain the product
$b_{\rm g} \sigma_8$, so we can easily scale the result for any preferred value
of $\sigma_8$.

In principle one should marginalise over the prior associated with each
parameter. However, as we show in Section~\ref{results}, the
cross-correlation signal is very weak and is actually compatible with the
null hypothesis of no correlation.  By widening the priors on $b_{\rm g}$
and $\sigma_8$, we find this makes the signal even less significant.
We are aware that widening priors for each parameter will reduce the
significance of our result, but we chose to opt for a more
optimistic view, and acknowledge that the significance of our result is
tentative.

\section{Galaxy Biasing from the Galaxy ACF}\label{bias}

\begin{figure}
\centerline{\includegraphics[width=10cm,bb=0 270 600 600,
clip]{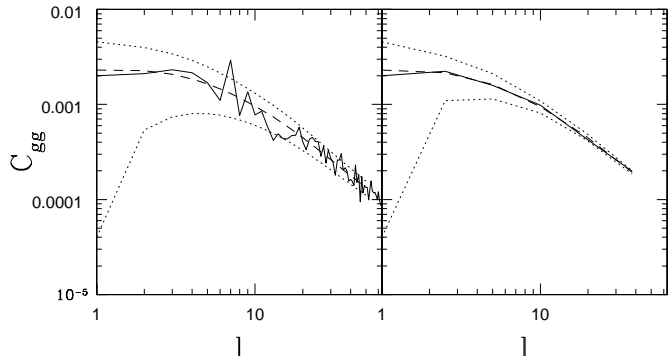}}\caption{The raw (left) and binned
(right) 2MASS angular power spectrum (solid line) for galaxies
$12<K_{20}<14$, with theory (dashed) for the best fit bias
$b_{g}=1.40$, and 1$\sigma$ gaussian error bars (dotted). The
expected poisson noise has been subtracted from the data.  We use
$\Omega_m=0.30$, $\Omega_b=0.05$, $\sigma_8=0.75$, $h=0.7$ as our
fiducial model.  } \label{figCgg}\end{figure}

We compute the ACF of 2MASS, $C_{gg}(\ell)$, and use it to constrain the
galaxy bias, $b_{\rm g}$, in Equations~\ref{CggEx} (and \ref{CggSA}).
Since we assume a constant bias across the magnitude shells we
make the fit using all the galaxies at once, $12 < K_{20} <
14$. The redshift distribution for all of 2MASS is plotted in
Figure~\ref{figdNdz}, and in Figure~\ref{figCgg} we plot the
results of the bias fitting.

Rigorously, Equations~\ref{CgtEx} and~\ref{CggEx} (and
~\ref{CgtSA} and~\ref{CggSA}) only hold for a \emph{linear} matter
power spectrum (calculated using CAMB ~\citep{Lewis:1999bs}), 
as we have assumed the redshift dependent power
spectrum was separable: $P(k,z)=D^2(z)P_{lin}(k)$. We observe that using
a linear power spectrum, one can fit the ACF well for linear scales,
i.e. for $\ell < 30$. However we find that using a
non-linear power spectrum (from ~\cite{Smith:2003p}, as implemented 
in CAMB) provides a better fit
to the data, and to higher $\ell$, suggesting that using
$P(k,z)=D^2(z)P_{non-lin}(k)$ can be considered a valid
approximation to about $\ell \sim 50$; since the bias does not change on
the depth of our survey, we do not expect its scale-dependence to
change much.  This issue does not arise when considering the
CCF, as the ISW effect should only arise on linear scales.

We therefore decide to fit the ACF using a non-linear power spectrum for 
$\ell\le 50$ (to avoid
highly non-linear scales), and we bin the data into 6
logarithmically spaced bins. This reduces correlation between
different multipoles so that we can assume the bins are independent
and have gaussian error bars. Hence, scatter about the expected
signal is just due to cosmic variance, and the likelihood can be
written: \be -2\ln {\mathcal L} = \det({\rm M}) + {\bf d}^T{\rm
M}^{-1}{\bf d} + const. \label{like}\ee Where the diagonal terms
of $\rm M$ contain the variance of $C_{gg}$ (e.g.: \cite{Dodelson:2003}), \be
\sigma^2(C_{gg})\approx \frac{1}{f_{\rm sky}}\frac{2}{2\ell+1}\:C^2_{gg}\label{fsky}\ee
and ${\bf d}=(\hat{C}_{gg}-C_{gg})$; $f_{\rm sky}$ is the fraction of sky
observed; $\hat{C}_{gg}$ is our observed overdensity ACF after shot
noise substraction, given by $1/\bar N$ where $\bar N$ is the mean number of
galaxies per steradian; $C_{gg}$ is the
theory (Equation~\ref{CggEx} and \ref{CggSA}). Note that the covariance matrix
${\rm M}$, and its determinant, depend on $b_{\rm g}$.

\begin{figure}
\centerline{\includegraphics[width=8cm,bb=60 150 620 500]{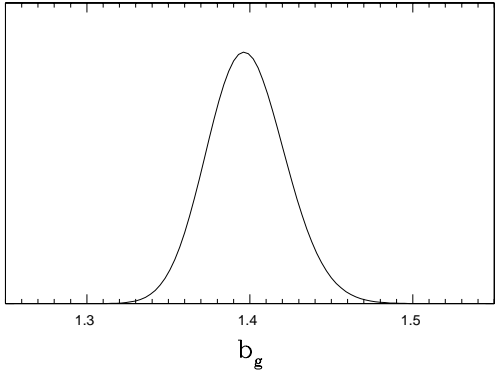}} 
\caption{The (unnormalized) likelihood of $b_{g}$
from fitting to the 2MASS galaxy auto-correlation function for our flat fiducial 
cosmology ($\Omega_m =
0.30$, $\Omega_b = 0.05$, $h=0.7$, $\sigma_8=0.75$). The best fit bias
is $b_{\rm g}=1.40 \pm 0.03$ from fitting for the multipole range
    $\ell=1-50$.  When fitting over a strict linear range
    ($\ell=1-25$) the best fit bias changes slightly to $b_{\rm
    g,lin}=1.38 \pm 0.05$ which does not affect the $\chi^2$ values
    reported in Table \ref{Chi}.  The auto-correlation $C_{gg}$  is
    shown in Figure~\ref{figCgg} with the fit $b_{\rm g}=1.40$ .}


\label{biaslike}\end{figure}

We fit for the bias using our fiducial model (flat universe with
$\Omega_m=0.30$, $\Omega_b = 0.05$, $h=0.7$, $\sigma_8=0.75$) and we
plot the resulting likelihood curve in Figure~\ref{biaslike}. The best
fit value for the multipole range $\ell=1-50$ is $b_{\rm g}=1.40 \pm 0.03$ 
(to 2 d.p) at 1$\sigma$ (fitting for
a gaussian).  
If we remove non-linear scales to the bias fitting, and consider only
the multipole range $\ell=1-25$, the error on the bias increases slightly while
its value remains roughly the same ($b_{\rm g,{\rm lin}}=1.38 \pm 0.05$).
We find our results in Section~\ref{results} do not differ significantly for
these two values, and herein we take $b_{\rm g}=1.40 \pm 0.03$.

The determination of the bias is particularly sensitive to
$\sigma_8$ because they both act as overall normalization factors, and
in linear theory $C_{gg}\propto (b_{\rm g}\sigma_8)^2$.

\section{Cross-Correlation Method}\label{method}

We perform the cross-correlation in harmonic space. We have used
masks on all the maps, and thus we cannot obtain true values of
the multipole coefficients $a_{\ell m}$, as the power of the harmonic estimator
will be reduced and correlations will be
induced between multipoles. However, we scale for the loss of sky
cover and use a full covariance matrix to account for the
correlation between bins. We have obtained the spherical harmonic
coefficients of our four 2MASS $K_{20}$ shells, and our four WMAP3
maps, as outlined in Section~\ref{data}. We perform the
cross-correlation:

\be C_{gT}(\ell)=\frac{1}{(2\ell+1)}\sum_{m}Re(a^{g}_{\ell
m}a^{T*}_{\ell m})\ee

\begin{figure}\centering
\psfig{file=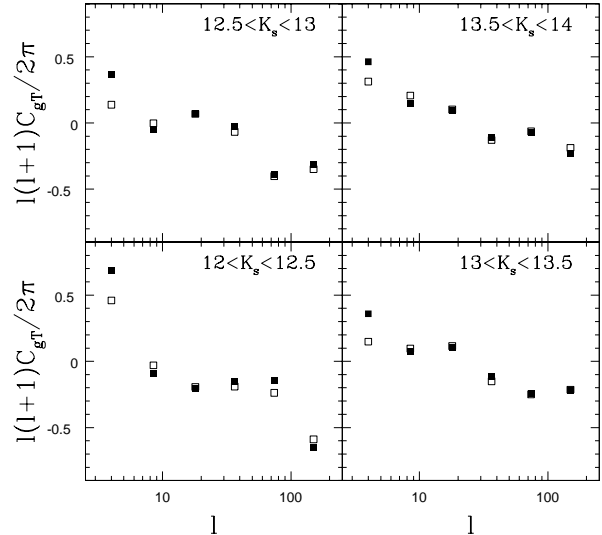,width=8cm}\caption{The
cross-correlation $C_{gT}$ ($\mu$K) of galaxies from the 2MASS
catalogue, and the CMB from the V-band WMAP1 (open squares) and WMAP3 
(filled squares) results. See Figure~\ref{3yr} for
error bars.  The power in the WMAP3
maps have changed very little (see for example Figure 19
in Hinshaw 2006), yet we see a slight change in the cross-correlations, especially in the
first point which corresponds to $\ell=3-5$. The difference may be due to a
slight change in the structure of these multipoles, perhaps due to the
improved gain model (Jarosik 2006).}\label{years}\end{figure}

We further bin the data, using logarithmically spaced bins and
$\ell \ge 3$. We avoid $\ell=2$ due to its anomalously low power
in the CMB. For the analysis we use 5 bins $3\le \ell \le 30$, as
this is where the ISW signal is expected to dominate. In
Figures~\ref{years} and ~\ref{3yr} we
plot the correlation using 6 bins with $3 \le \ell \le 200$.

In Figure~\ref{years} we compare the $C_{gT}$ results of the WMAP1
and WMAP3 V-band data. Surprisingly we see a slight
change, especially in the first point which corresponds to
$\ell=3-5$. The power in the WMAP3 maps has changed very little
(see for example Figure 19 in~\cite{Hinshaw:2006ia}), and thus the
difference must be due to a slight change in the structure of
these multipoles, perhaps due to the improved gain
model~\citep{Jarosik:2006ib}. Note that although the reported
power spectrum has changed at low-$\ell$ this is due to change in
the likelihood analysis, rather than a change in the underlying data ~\citep{Hinshaw:2006ia}.

We will be comparing two hypotheses: a null hypothesis of no
cross-correlation, and that returned by Equation~\ref{CgtEx}
(and~\ref{CgtSA}) for our fiducial $\Lambda \rm CDM$ cosmology. To assess the
fit we will compare the $\chi^2$ values returned by the two
hypotheses.  Similarly we can compare the unnormalized
evidence ${\mathcal E}$, where $-2\ln{\mathcal E} = \chi^2$ (if all
parameters are fixed), or given by Equation~\ref{Ev} (if we marginalize
over a parameter, where $b_{\rm g}$ in the equation can be replaced by
any parameter). For this we use a covariance matrix, estimated from
simulations, and thus we account for the correlation between
$\ell$ bins and those between $K$-shells.

We make 500 simulations of a Gaussian CMB using the best fit
theoretical $C_{TT}$ from WMAP3. We apply the Kp2 mask,
and correlate them with the four $K$-shell 2MASS maps. By not
varying the 2MASS maps we have slightly underestimated the errors,
because we have not accounted for the cosmic variance of the 2MASS
data.   We also note that the ISW signal in not inbuilt in the
simulations; this should be inconsequential since this signal is negligible compared to the cosmic
variance of the CMB.

We compute the $\chi^2$ to find the favoured model and the
improvement $\Delta(-2\ln {\mathcal E})=\Delta(\chi^2)$.  As usual we have
$\chi^2={\bf d}^TM^{-1}{\bf d}$, ${\bf d}=(\hat{C}_{gT}-C_{gT})$,
where $C_{gT}$ is calculated for the two models, one with no
correlation, i.e. $C_{gT}=0$ and another with correlation due to ISW
effect, given by equation ~\ref{CgtEx} (and~\ref{CgtSA}) for our
fiducial cosmology.  The covariance matrix is defined as \be \displaystyle{M_{ij}\equiv
  \left< (\rm
d_i-\left<d_i\right>)(d_j-\left<d_j\right>)\right>}\ee and is calculated
from simulations in which the galaxies and the CMB are uncorrelated.
We use 5 logarithmically spaced bins at low-$\ell$ 
where the expected signal dominates ($\ell=3-30$), and we include all four 
$K_{20}$ shells in the analysis, thus
$i,j=1,..,20$.  The fact that cosmic variance of 2MASS is not included suggests
our simulations will \emph{underestimate} the errors.  We perform a 
consistency check on our covariance matrix by analytically estimating
the diagonal elements of the covariance matrix:
\be M_{ii}=\sigma^2(C_{gT})=\frac{1}{f_{sky}(2\ell+1)}\big(C_{gT}^2+C_{gg}C_{TT}\big)\label{genfsky}\ee
which is the general form of Equation \ref{fsky}.  We find that error
bars calculated using Equation~\ref{genfsky} are larger but of the
same order of magnitude than those estimated from simulations.

\section{Cross-Correlation Results}\label{results}

\begin{table*}
\caption{Log evidence ($-2\ln{\mathcal E}$) values for cross-correlation 
of each WMAP3 maps (V, W, Q and ILC) with four 2MASS magnitude shells, 
using different model assumptions and priors.}
  \begin{tabular}{lccccc}
  \hline
&1&2&3&4&5\\
  \hline
  & Null Hypothesis & $\Lambda \rm CDM$ &$\Lambda$CDM &
  $\Lambda$CDM& $\Lambda$CDM\\
  & of no correlation& $b_{\rm g}=1.40$ & prior on $b_{\rm g}$&
  $\Omega_\Lambda=0.85$& marginalized \\
 &&&&& over $\Omega_\Lambda$\\

 \hline
$ILC$ & 11.3 & 9.7 &9.7&7.3&10.1\\
$Q$ & 12.1 & 10.4 &10.4 &8.1&10.9\\
$V$ & 11.0 & 9.5 &9.5&7.4&10.0\\
$W$ & 10.8 & 9.1 &9.1&6.9&9.6\\
\hline
\end{tabular}\\
\medskip
\medskip
\raggedright 
Log evidence ($-2\ln{\mathcal E}$) values for cross-correlation of
each WMAP3 maps (V, W, Q and ILC) with four 2MASS magnitude shells,
using different model assumptions and priors.  In all cases $\sigma_8$
is taken to be $0.75$.  {\bf Model 1} is the null hypothesis of no
correlation.  In {\bf model 2}, we have considered a flat universe
with $\Omega_m = 0.30$, $\Omega_b = 0.05$, $h=0.7$ and $b_{\rm
  g}=1.40$ (for $b_{\rm g}=1.38$ only the W map result changes from
9.1 to 9.2).   In {\bf model 3}, we have widened the prior  used on
$b_{\rm g}$ to that which we obtained by fitting the 2MASS ACF on
scales $\ell=1-50$, i.e. $b_{\rm g}=1.40\pm 0.03$ (the results are
unchanged when using $b_{\rm g,lin}=1.38\pm 0.05$, the bias obtained
from linear scales only, i.e., $\ell=1-25$).   {\bf Model 4} is best
fit $\Omega_\Lambda = 0.85$ in a universe with no curvature.  In {\bf
  model 5}, we have marginalized over $\Omega_\Lambda$ using the
likelihood in Figure ~\ref{OmL} assuming a flat geometry and a uniform
prior on $\Omega_\Lambda$ in the range $[0;0.95]$.  The evidence ratio 
for model 1 (null hypothesis) and 5
(marginlized $\Lambda$CDM) is $\Delta(-2\ln{\mathcal E})\approx 1.1$
which means the data prefer a $\Lambda$CDM cosmology to the null
hypothesis, but only marginally.  For models 1, 2 and 4 the evidence
is related to the $\chi^2$ by $-2\ln{\mathcal E}=\chi^2$.  
\label{Chi}
\end{table*}

In this section we discuss the significance of the cross-correlation
results and determine an upper limit on $\Omega_\Lambda$.

\subsection{Null Hypothesis}

\begin{figure}\centering
\psfig{file=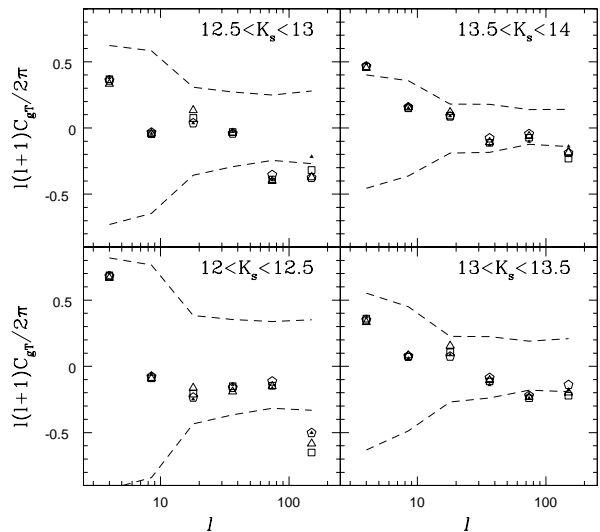,width=8cm}\caption{Cross-correlation
$C_{gT}(\mu K)$ results for the ILC (small triangle) , Q (open
triangle), V (open square), and W (open pentagon) CMB maps with
different magnitude bins of the 2MASS galaxy surveys.  The dashed
lines are 1$\sigma$ error bars about the null hypothesis, as evaluated from
simulations. An ISW effect is expected to be \emph{achromatic}, which
is what we observe, but the
null hypothesis is not ruled out.}\label{3yr}\end{figure}

In Figure~\ref{3yr} we plot the results of the cross-correlation, for
6 logarithmically separated bins between $\ell$=3-200,
with 1$\sigma$ error bars.  The cross-correlation is
\emph{achromatic}, indicative of an ISW type cross-correlation. However the
results appear completely compatible with the null hypothesis,
and in fact scatter much less than expected.

We use five logarithmically separated bins between $\ell$=3-30 to
compute the $-2\ln{\mathcal E}$ values for the (ILC,Q,V,W) maps.  These
values can be found in Table~\ref{Chi} (model 1). By comparing to 
simulations, we find these values are low at the
$\sim95\%$ level.

\subsection{$\Lambda \rm CDM$ Fiducial Model}
 To rule out the null hypothesis, at face value
without considering a competing theory, we would actually need a
high $\chi^2$ value. However, we are comparing two theories and
thus can ask the more subtle question of which the data prefer.  To
do this we can consider the ratio of the evidences,
$\Delta(-2\ln{\mathcal E})$.

\begin{figure}
\centerline{ \includegraphics[bb=20 160 350
680,clip,width=6cm]{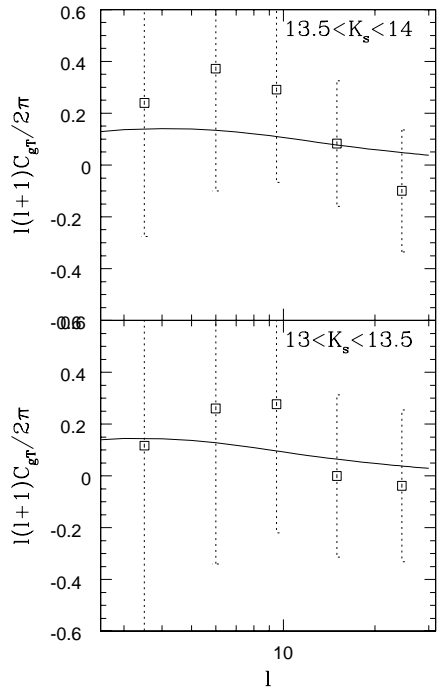}}\caption{The observed
$C_{gT}$ ($\mu$K) (\emph{squares}) from the V-band CMB map and the
furthest magnitude shells of the 2MASS Survey ($13<K_{20}<13.5$ and
$13.5<K_{20}<14$), with 1$\sigma$ cosmic variance.  These two shells 
contain the most galaxies.  The
solid line represents the theory for our fiducial model (flat universe
with $\Omega_m=0.30$, $\Omega_b = 0.05$, $h=0.7$, $\sigma_8=0.75$) and
$b_{\rm g}=1.40$. We have used 5 bins between
$\ell=3-30$ which corresponds to scales for which the ISW signal
dominates.}\label{3yr2}\end{figure}

In Figure~\ref{3yr2} we plot the results again for the two furthest
redshift shells, which contain the most galaxies, for 5 bins at 
low-$\ell$ where the expected signal dominates. We also show the 
theoretical signal expected from our fiducial $\Lambda$CDM.

Using these bins and the four $K_{20}$ shells, 
the $-2\ln{\mathcal E}$ values found for the ISW
theory, using $b_{\rm g}=1.40$ and
our fixed fiducial cosmology are reported in Table~\ref{Chi} (model 2). 
There is an improvement of
$\Delta (-2\ln {\mathcal E}) \approx1.5$, and thus we confirm that the 
data prefer the
ISW theory to the null hypothesis (of no correlation, or equivalently 
Einstein-de Sitter). However rule of thumb has it
that `strong' evidence is $\Delta(-\ln {\mathcal E})\ge 3$, and
thus this improvement is not compelling.


We widen the uncertainty around the bias value, using the error bars
from Section~\ref{bias}.  There we found $P(b_{\rm g})$ was
approximately a gaussian with $(\mu, \sigma) = (1.40, 0.03)$,
although we use the curve from Figure~\ref{biaslike} for our
marginalization:

\be {\mathcal E}=\int P({\rm data}|{\rm theory},b_{\rm g})P(b_{\rm
  g})\rd b_{\rm g}\label{Ev}\ee
where we have assumed uniform priors on the theory. In Table
  ~\ref{Chi}, we record our $-2\ln{\mathcal E}$ results for a $\Lambda$CDM with a prior on
  the bias (models 3), which are identical at 2 d.p. with those
  obtained from fixing $b_{\rm g}=1.40$.  We get very similar results if 
we use the 2D probability distribution function
  $P(b_{\rm g},\sigma_8)$ and marginalize over both $b_{\rm g}$ and
  $\sigma_8$ (using a gaussian with $\mu_{\sigma^2_8}=0.56$ and 
$\sigma_{\sigma^2_8}=0.08$).  The results are also unchanged if we use the strictly
  linear bias $b_{\rm g, lin}= 1.38 \pm 0.05$.

\subsection{Assessing the goodness of fit}

By comparing our $\chi^2$ values to those obtained from simulations, we
find that they are low to $\sim 95\%$, which is also evident from the lack 
of scatter in Figure~\ref{3yr}.
However, we are interested in maximising the evidence, or
equivalently minimising the $\chi^2$, and thus we find that moving
from the null hypothesis to the $\Lambda$CDM model, the fit is
improved (raising the interesting question: is minimising the $\chi^2$ always
appropriate?). 


In the above analysis, we consider 20 correlated data points (five
angular points in each of the four radial shells), and
calculate the exact $\chi^2$ using a full covariance matrix.  We can also consider each
magnitude shell separately and calculate the $\chi^2$ for
each one, as a consistency check.  
The $\chi^2$ values obtained for the null hypothesis are (1.70, 0.67,
0.94, 2.36) going from the nearest to the furthest $K_{20}$ 
shell, and for our fiducial model we find (1.72, 0.58, 0.83, 2.16).
When the data are
thus considered, our fiducial model is not always a better fit than
the null hypothesis, and data in the closest shell prefer the null
hypothesis.



\subsection{Upper Limit on $\Omega_\Lambda$ and Marginalization}

Above we have compared our fiducial cosmology to the null hypothesis. Alternatively we can chose to consider only our fiducial $\Lambda \rm CDM$ model and use
the cross-correlation to constrain its parameters.   We vary 
$\Omega_\Lambda$, and $\Omega_m$ keeping all
other parameters fixed ($\Omega_m=1-\Omega_\Lambda$), and fit it to
the measured correlation.  

\begin{figure}\centering
\includegraphics[bb=60 150 620 500 ,width=8cm]{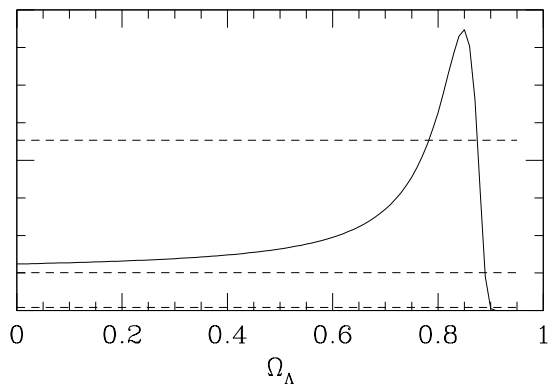}
\caption{The (unnormalized) likelihood of $\Omega_\Lambda$ from fitting to the
observed $C_{gT}$ results (V band). The maximum is at $\Omega_\Lambda=0.85$
Horizontal lines indicate the 1,2,3 $\sigma$ levels, equivalent to
the decrease in the likelihood value. We find the corresponding
upper limits on $\Omega_\Lambda$ at 0.87, 0.89,0.90.  Corresponding
$\chi^2$ values can be found in Table~\ref{Chi}.} \label{OmL}\end{figure}

In Figure~\ref{OmL} we plot the resulting likelihood. We find a best fit of
$\Omega_\Lambda=0.85$, and upper limits of 0.87, 0.89, 0.90 at 1,
2, 3$\sigma$ respectively. Corresponding $-2\ln{\mathcal E}$ 
values can be found in Table~\ref{Chi} (model 4).

This relatively high
value for $\Omega_\Lambda$ is in good agreement with other studies
of the ISW effect;~\cite{Cabre:2006qm} find
$\Omega_\Lambda=0.8-0.85$ at the 1$\sigma$ level.
As can be seen, the null hypothesis ($\Omega_\Lambda=0$) is 
less than 2$\sigma$ away from the best fit result, confirming that 
we cannot confidently rule it out.

In the last column of Table~\ref{Chi}, we present the marginalized
evidence for our $\Lambda$CDM model (model 5), where the prior on the theory is
flat over $\Omega_\Lambda = 0 - 0.95$, and the likelihood of our model is taken
from Figure~\ref{OmL}. For this model $\Delta(-2\ln{\mathcal E}) \approx
1.1$ on average, so the data prefers a $\Lambda$CDM model
to the null hypothesis, but only marginally.

Bearing in mind this issue, and the fact that the $\chi^2$ values are
low for all models, any claim for an ISW detection using 2MASS and WMAP3
remains \emph{tentative}, and indeed our results are also consistent
with the null hypothesis of no correlation within the $2 \sigma$ level.

\section{Axis of Evil}\label{axis} We briefly consider other possible
sources and statistics of a
positive cross-correlation between the CMB and LSS. As discussed in 
Section~\ref{intro}, interesting
anomalies dubbed the ``Axis of Evil'' (AoE) have been observed in
the WMAP data of the CMB that indicate a possible departure from
statistical isotropy~\citep{Land:2005ad}. Here we give a brief description 
of the AoE anomaly, but the reader should
refer to \cite{Land:2005ad} for a detailed description.  

In harmonic
space, a statistically isotropic map is expected to have independent $\ell$
modes.  For a given multipole $\ell$, the power should also be distributed
randomly amongst each $m$-mode, but it is possible to rotate any map so
that the power in a given multipole is mostly in a
given $m$-mode.  However, if for a given frame, several multipoles
have power mostly distributed in one $m$-mode, then the map can no
longer be considered statiscally isotropic.  \cite{Land:2005ad} found
that there existed a set of nearly identical frames (with axis
dubbed the ``Axis of Evil'')  in which the multipoles of the WMAP1
maps showed phase correlations.

Possible sources of these anomalies are: foreground contamination;
astrophysical effects ({\it e.g.}, lensing and moving cluster effect);
alternative cosmological paradigms. If the AoE was in some way due
to local inhomogeneities ({\it e.g.},~\cite{Vale:2005mt}) then one
might also expect a $C_{gT}$ cross-correlation on these large
scales.  We investigate the AoE anomaly in two ways.  Firstly, we
rotate the 2MASS data to the frame of the WMAP AoE frame and search
for anomalous phase correlations in 2MASS in that frame.  Secondly, we
rotate the 2MASS data in all possible directions, searching for an AoE
type anomaly in all frames.

\begin{figure}
\centerline{\includegraphics[width=8cm,bb=50 180 600
550,clip]{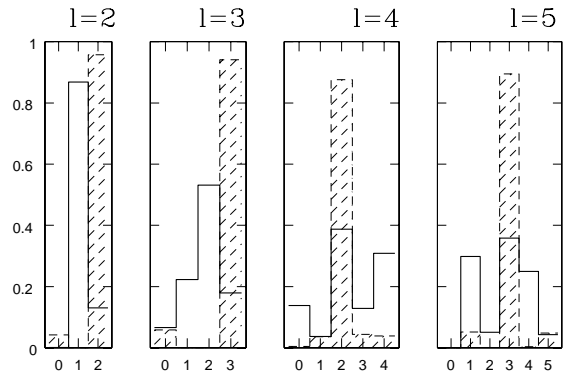}} \caption{The fraction of power in each
$m$-mode for multipoles $\ell=2,3,4,5$. The $a_{\ell m}$s are
computed in the Axis of Evil (AoE) frames returned by the CMB. By
definition the multipoles of the CMB are dominated by one $m$-mode
in these frames (shaded). We do not observe a similar pattern in
the 2MASS results (solid line).  Negative values of $m$ are not shown
as $a_{\ell -m}=a^*_{\ell m}$. }\label{aoe}\end{figure}

For the first test, we examine if the AoE signal observed in the CMB is also present
in the 2MASS data. In Figure~\ref{aoe} we plot the power observed
in each $m$-mode, for the AoE frames returned by the CMB. For each multipole, this is
the frame where one $m$-mode dominates. We plot the power ratio
$2|a_{\ell m}|^2/C_\ell$ (without the 2 factor for $m=0$)
evaluated in the AoE frames, for the 2MASS ($12<K_{20}<14$) as
well as the CMB (using the cleaned ILC map
of~\cite{Tegmark:2003ve}). By definition, in these frames the CMB
observes a pattern of one $m$ significantly dominating each
multipole. We do not observe a similar pattern in the 2MASS data,
and thus conclude that if there exists a source responsible for the AoE
features, then is is unlikely to be the same source of the
$C_{gT}$ correlation observed above.

For the second test, we looked at 2MASS independently over $\ell=2-20$ and found no
evidence for any AoE type structure (a correlation of
multipole frames as defined above). Clearly the 2MASS catalogue is
highly anisotropic, with structures such as the Supergalactic
Plane visible by eye as well as the Galactic mask. Is it not
strange that a statistic that measures evidence for deviations
from statistical isotropy in the CMB does not find any such
feature in 2MASS? The AoE statistic highlights a very particular
type of phase correlation, in harmonic space, and is by no means a
conclusive test for general deviations from statistical isotropy.
The fact that a clearly non-gaussian and anisotropic map, such as
2MASS, does not return a anomalous AoE signal for $\ell=2-20$
perhaps highlights the weakness of this statistic, and throws
caution at how one defines and selects appropriate statistics.
Perhaps the non-gaussianity is `washed out' by the depth of the
survey - anomalies might be more significant in the shallowest part of
the survey.
However, there is currently no conclusive way to test a map for
deviations from statistical isotropy.


\section{DISCUSSION}\label{discuss}

We calculate the cross-correlation between the 2MASS galaxy
survey and the WMAP3 data. This updates the work
of~\cite{Afshordi:2003xu} who cross-correlated 2MASS with WMAP1.

The cross-correlation signal expected in a $\Lambda$CDM Universe
scales with the linear galaxy bias, $b_{\rm g}$, and in linear theory
with the product ($b_{\rm g}\sigma_8$).  We fix $\sigma_8 =
0.75$, and use a flat fiducial cosmology (based on inflation).  We
use $\Omega_m = 0.30$ (based on supernov{\ae} Ia data), $\Omega_b =
0.05$ (based on Big Bang nucleosynthesis), $h=0.7$ (based on the HST
key project), $n = 1$ (i.e., the Harrison-Zeldovich spectral index).

Fitting the fiducial cosmology to the angular auto-correlation
function of the 2MASS galaxy survey for $12<K_{20}<14$, yields a
linear bias value of $b_{\rm g}=1.40\pm 0.03$, assessed from multipole
scales $\ell=1-50$ ($b_{\rm g,lin} = 1.38 \pm 0.05$ assessed from
  multipole scales $\ell=1-25$).  As the 2MASS galaxy survey is
shallow, we assume $b_{\rm g}$ is constant with redshift over the
depth of our survey.

The measured cross-correlations obtained from four different WMAP maps (V,
W, Q, and ILC) and four different magnitude shells of 2MASS
($12<K_{20}<12.5$, $12.5<K_{20}<13$, $13<K_{20}<13.5$, and
$13.5<K_{20}<14$) show an achromatic signal, as expected from an ISW
effect.
However, the observed signal is also within the 1$\sigma$ error bars
obtained from cross-correlating random simulations of the CMB with
2MASS data.  This means the data are consistent with the null hypothesis of no correlation.

We compare our observation with the ISW signal expected in our
fiducial $\Lambda$CDM model, with a fixed bias and an uncertaintly around the bias
value which we found when fitting the auto-correlation function. The 
$\Lambda$CDM model finds a lower chi-squared, and thus is a better fit
compared to the null hypothesis, but 
in both cases of treating the bias the evidence change is only
$\Delta(-2\ln{\mathcal E})\approx 1.5$.
Whichever model is considered - the null hypothesis or a $\Lambda$CDM
universe - the $\chi^2$ values are low when compared to simulations.

Varying the Dark Energy density component of the $\Lambda$CDM model 
and assuming flatness, we find that the data
prefer a high value of $\Omega_\Lambda = 0.85$, with
$\Omega_\Lambda<0.89$ (95\% CL).

Thus, there is a higher correlation between WMAP3 and 2MASS than that
expected by a $\Lambda$CDM universe with $\Omega_\Lambda = 0.7$.
However, the observed cross-correlation may not only be due to an ISW
effect.  Other signals might contribute to the correlation: positive
curvature (even if very small) or cosmic magnification (higher redshift
galaxies can contribute more than expected if they are lensed by the
lower redshift galaxies and their luminosity function is such that the
effect leads to a positive correlation).  

When we marginalize over $\Omega_\Lambda$, the evidence ratio between
the null hypothesis and $\Lambda$CDM becomes $\Delta(-2\ln{\mathcal E})
\approx 1.1$ so we can say that the data prefer a
$\Lambda$CDM universe, but only marginally. In any case, the correlation observed is
consistent with both hypotheses within 2$\sigma$.  

We also investigate ``Axis of Evil'' (AoE) type anomalies, which detect phase
correlations between different multipoles in harmonic space.  These
phase correlations are not expected in a statistically isotropic map ~\citep{Land:2005ad}.
We do not observe an AoE type of structure in the 2MASS
catalogue. As non-Gaussian features are expected in the LSS we 
feel this result raises issues about the
use of the AoE statistic as a general test for statistic anisotropy.

We do not observe correlations between the CMB fractional
power distribution as measured in its AoE frame and that of 2MASS,
constraining the possible explanations of the low $\ell$ anomalies in
the CMB.

Future spectroscopic and photometric galaxy redshift surveys
(\emph{e.g.}, the Dark Energy Survey, the Wide Field Multi-Object
Spectrograph) will yield more galaxies out to higher
redshifts.  \cite{Afshordi:2004kz} showed that an all sky survey with
10 million galaxies and uniform sky coverage between $0<z<1$ would
lead to a detection of the ISW effect at the $5\sigma$ level.  
It remains to be assessed what combination of depth and
sky coverage is optimal for detecting the ISW effect.

We highlight that a claim for an ISW detection could be greatly
weakened if one considers some of the uncertainty around the
cosmological parameters. For our 2MASS-WMAP3 correlation the
signal-to-noise ratio is poor, however with a stronger data set a
full Markov Chain Monte Carlo exploration of parameter space should be
done.

\section*{Acknowledgments}
We thank Jo\~{a}o Magueijo, Niayesh Afshordi,
Sarah Bridle, Chris Blake, Carlo Contaldi, Jochen Weller and Saleem Zaroubi for useful
discussions and assistance. AR also thanks PLR. AR is supported by
the Perren Fund. KRL is funded by PPARC. OL acknowledges a PPARC Senior
Fellowship.  We used HEALPix
software~\cite{Gorski:1998vw}, the 2MASS
catalogue\footnote{http://www.ipac.caltech.edu/2mass/}, the WMAP
data\footnote{http://map.gsfc.nasa.gov/}, and the Galaxy
extinction maps of~\cite{Schlegel:1997yv}.

Results were computed on the UK-CCC COSMOS supercomputer which is
supported by HEFCE and PPARC, and in cooperation with SGI/Intel
utilising the Altix 3700 supercomputer. We thank Stuart Rankin for
COSMOS assistance and other helpful communications.

\bibliographystyle{mn2e}
\bibliography{ISW_FINAL}

\bsp

\label{lastpage}

\end{document}